\documentclass[twocolumn,showpacs,preprintnumbers,amsmath,amssymb,superscriptaddress,pra]{revtex4}

\usepackage{graphicx}
\usepackage{dcolumn}
\usepackage{bm}

\begin{document}

\preprint{APS/123-QED}

\title{Analysis of a Single Atom Dipole Trap}

\author{Markus Weber} \email{markus.weber@physik.uni-muenchen.de}
\affiliation{Department f\"ur Physik, Ludwig-Maximilians-Universit\"at
M\"unchen , D-80799 M\"unchen, Germany}

\author{J\"urgen Volz} 
\affiliation{Department f\"ur Physik, Ludwig-Maximilians-Universit\"at
M\"unchen , D-80799 M\"unchen, Germany}

\author{Karen Saucke} \affiliation{Department f\"ur Physik,
Ludwig-Maximilians-Universit\"at M\"unchen , D-80799 M\"unchen, Germany}

\author{Christian Kurtsiefer} \affiliation{Department of Physics, National
University of Singapore, Singapore}

\author{Harald Weinfurter} \affiliation{Department f\"ur Physik,
Ludwig-Maximilians-Universit\"at M\"unchen , D-80799 M\"unchen, Germany}
\affiliation{Max-Planck-Institut f\"ur Quantenoptik, 85748 Garching, Germany}

\date{\today}

\begin{abstract}
We describe a simple experimental technique which allows to store a single
$^{87}$Rb atom in an optical dipole trap. Due to light-induced two-body
collisions during the loading stage of the trap the maximum number of captured
atoms is locked to one. This collisional blockade effect is confirmed by the
observation of photon anti-bunching in the detected fluorescence light. The
spectral properties of single photons emitted by the atom were studied with a
narrow-band scanning cavity. We find that the atomic fluorescence spectrum is
dominated by the spectral width of the exciting laser light field. In addition
we observe a spectral broadening of the atomic fluorescence light due to the
Doppler effect. This allows us to determine the mean kinetic energy of the
trapped atom corresponding to a temperature of 105 $\mu$K. This simple
single-atom trap is the key element for the generation of atom-photon
entanglement required for future applications in quantum communication and a
first loophole-free test of Bell's inequality.
\end{abstract}

\pacs{03.67.-a, 32.80.Pj, 42.50.Vk, 42.50.Ar}
\maketitle

\section{Introduction}

The coherent control of a single quantum emitter is a crucial element for the
effective generation of single photons and even more for the generation of
entanglement between the emitted photon and the radiating quantum system
\cite{Karen,Simon03,Blinov04,Kuzmich04,Weber05}. It is thus of fundamental
importance for future applications in quantum communication and information
processing, like quantum networks \cite{Nielsen} or the quantum repeater
\cite{Briegel98}. So far, a great variety of possible experimental schemes has
been worked out, based upon the control of fluorescence from different kind of
emitters, like ions \cite{Neuhauser80,Walther87}, atoms
\cite{Schlosser01,Rempe02}, molecules \cite{Brunel99,Lounis00}, color centers
\cite{Kurtsiefer00,Brouri00} or semiconductor structures
\cite{Michler00,Santori01}.

Cold atoms - isolated from the environment by the use of standard laser
cooling and trapping techniques - are outstanding candidates for future
applications in quantum networking. On the one hand, narrow-band atomic
transitions can be used for the generation of single photons. On the other
hand, due to the intrinsic clarity of the well defined internal level
structure, atoms are also well suited for the realization of a long-lived
quantum memory. In particular, a single laser cooled $^{87}$Rb atom, localized
in a far-off-resonance optical dipole trap (FORT) \cite{Miller93,Schlosser01},
lends itself to store quantum information in the level structure of the atomic
ground state $5^2 S_{1/2}$ for a long time with a very small relaxation rate
\cite{Cline94}. The stored quantum information can in principle be converted
to flying qubits (photons) at a wavelength of 780 nm - suitable for low-loss
communication - and therefore be transmitted between specified remote
locations. And most important, the spontaneous decay of a single $^{87}$Rb
atom prepared in the $5^2 P_{3/2}, F'=0$ hyperfine level can be used to
generate entanglement between the spin state of the atom and the polarization
of the emitted photon \cite{Karen,Weber05} necessary for the scalable coupling
of quantum memories \cite{DiVicenzo00}.

In this paper, we report the observation and analysis of a single $^{87}$Rb
atom in an optical dipole trap that operates at a detuning of 61 nm from
atomic resonance. Atoms stored in this FORT have a very low photon scattering
rate and therefore negligible photon recoil heating. Confinement times up to 4
s are achieved with no additional cooling. Because of the small trap volume,
only a single atom can be loaded at a time \cite{Schlosser02}. To prove this
property the photon statistics of the detected fluorescence light has been
studied with an Hanbury-Brown-Twiss (HBT) setup. The measured second order
correlation function exhibits strong photon anti-bunching verifying the
presence of a single atom in the trap. In addition the two-photon correlations
show coherent dynamics of the population of the atomic hyperfine levels
involved in the excitation process. The observed correlations cannot be
explained by the simple model of a two-level atom \cite{Carmichael76}. In
order to simulate the second-order correlation function of the measured
fluorescence light, we numerically solve optical Bloch equations based on a
four-level model. Within experimental errors we find good agreement of the
theoretical predictions with the experimental data. Furthermore, we analyze
the spectral properties of the emitted photons with a scanning Fabry-Perot
interferometer (FPI). Due to the Doppler effect we observe a broadening of the
Rayleigh scattered atomic fluorescence spectrum relative to the spectral
distribution of the exciting laser light field. This broadening allows us to
determine the mean kinetic energy of the trapped atom corresponding to a
temperature of 105 $\mu$K.

\section{Experimental Setup}
 
\begin{figure}[t]
\includegraphics[width=8cm]{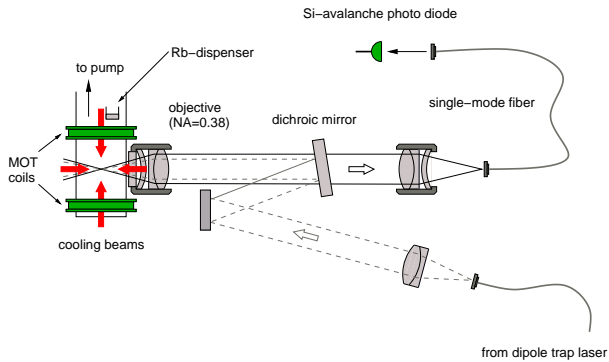}
\caption{(color online). Experimental setup of the dipole trap and
fluorescence detection: The dipole trap laser is focused at the intersection
of three pairs of counter-propagating laser beams for optical
cooling. Fluorescence light is collected with a microscope objective, coupled
into a single mode optical fiber, and detected with a silicon APD.}
\label{Bild:setup}
\end{figure}

In our experiment the FORT is generated by a Gaussian laser beam of a single
mode laser diode at a wavelength of 856 nm, which is focused down with a
microscope objective (located outside the vacuum chamber) to a waist of $3.5
\pm 0.2$ $\mu$m (see Fig. \ref{Bild:setup}). For a laser power of 44 mW we
calculate a trap depth of 1 mK and a photon scattering rate of 24 s$^{-1}$
\cite{Grimm99}. In order to load atoms into this FORT, we start with a cloud
of laser cooled atoms in a magneto-optical trap (MOT) \cite{Monroe90}. The MOT
is loaded from the thermal Rubidium background gas produced by a dispenser
operating slightly above threshold (residual gas pressure below 10$^{-10}$
mbar). This provides a macroscopic reservoir of cold atoms with a typical
temperature on the order of 100 $\mu$K. The dipole trap overlaps with the MOT
and thus by changing the magnetic field gradient of the MOT we can adjust the
loading rate of atoms into the dipole trap from 0.2 s$^{-1}$, without
quadrupole field, up to 1 atom per second at a magnetic field gradient of 1
G/cm. To assure optimal conditions for laser cooling in the dipole trap, the
magnetic field is compensated below a residual value of 300 mG by three
orthogonal pairs of Helmholtz coils generating a suitable bias field.

The fluorescence light scattered by atoms in the dipole trap region is
collected with the focusing objective and separated from the trapping beam
with a dichroic mirror. Then it is coupled into a single mode optical fiber
for spatial filtering and detected with a silicon avalanche photodiode
(APD). In this way it is possible to suppress stray light from specular
reflections of the cooling beams and fluorescence light from atoms outside the
dipole trap.

To load a single atom into the FORT, we switch on the cooling and repump laser
of the MOT and measure the fluorescence counting rate from the dipole trap. If
a cold atom enters the trap we observe an increase of the detected
fluorescence count rate. Typical photon counting rates are 500 - 1800 s$^{-1}$
per atom depending on the detuning and intensity of the cooling laser. From
the overlap of the detection beam (waist: $2.2 \pm 0.2$ $\mu$m) with the
emission characteristics of the emitted atomic fluorescence we calculate an
overall detection efficiency for single photons of 0.1 $\%$ including
transmission losses and the quantum efficiency of our Si-APD.

\begin{figure}[t]
\includegraphics[width=8cm]{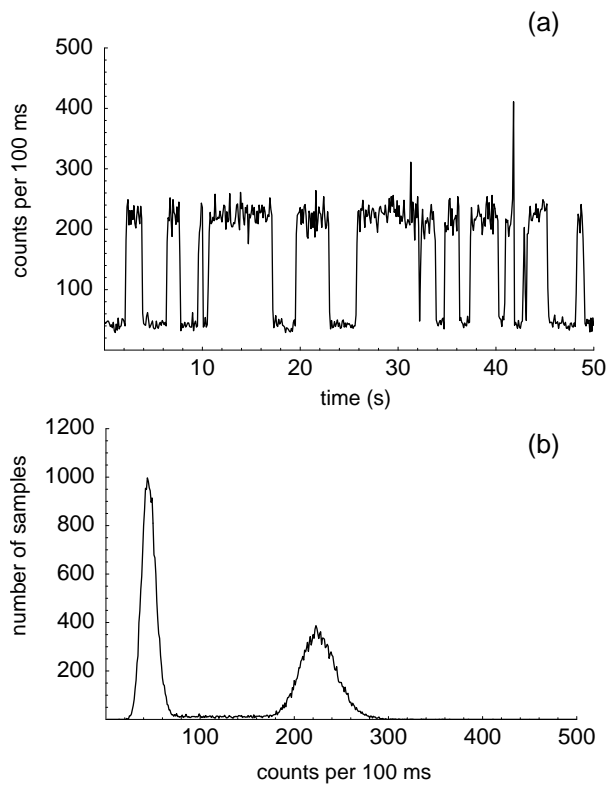}
\caption{(color online). Single atom detection. (a), number of photons counted
by an avalanche photodiode per 100 ms. (b), histogram of the photon-counting
data. Due to a collisional blockade effect \cite{Schlosser02,Weber05} only
counts corresponding to zero or one atom are observed.}
\label{Bild:singleatomtrace}
\end{figure}
 
The fluorescence rate exhibits the typical telegraph-signal structure (see
Fig. \ref{Bild:singleatomtrace}) jumping between background intensity (450
s$^{-1}$) when no atom is in the trap and a defined intensity level (2250
s$^{-1}$) corresponding to one atom. Other fluorescence intensities have
hardly been observed. This effect is caused by light-induced two-body
collisions, that together with the small dipole trap volume give rise to a
blockade mechanism which assures that only a single atom is trapped per
time. If a second atom enters the trap, inelastic two-body collisions
\cite{Wieman00} lead to an immediate loss of both atoms \cite{Schlosser02}. We
emphasize, that so far, a random telegraph signal was considered typical, when
the dark phase corresponds to the same atom as the bright one does, but with
the atom being shelved in a so-called dark state \cite{Bergquist86}. In
contrast, here a single atom enters and leaves the trap. From the fluorescence
trace in Fig. \ref{Bild:singleatomtrace} we determined a 1/e lifetime in
presence of cooling light of $2.2\pm0.2$ s. The mean lifetime without cooling
light is $4.4\pm0.2$ s. Due to interaction with the far-off resonant dipole
trap laser field, spontaneous Raman scattering leads to a change of the
population occupation of an atom initially pumped to the F=1 hyperfine ground
level. This hyperfine state changing scattering rate was determined in a
measurement, similar to \cite{Cline94}, to 0.1 s$^{-1}$ for a trap depth of
0.75 mK.

\section{Photon Statistics}

\begin{figure}[t]
\begin{center}
\setlength{\unitlength}{1cm} \includegraphics[width=8cm]{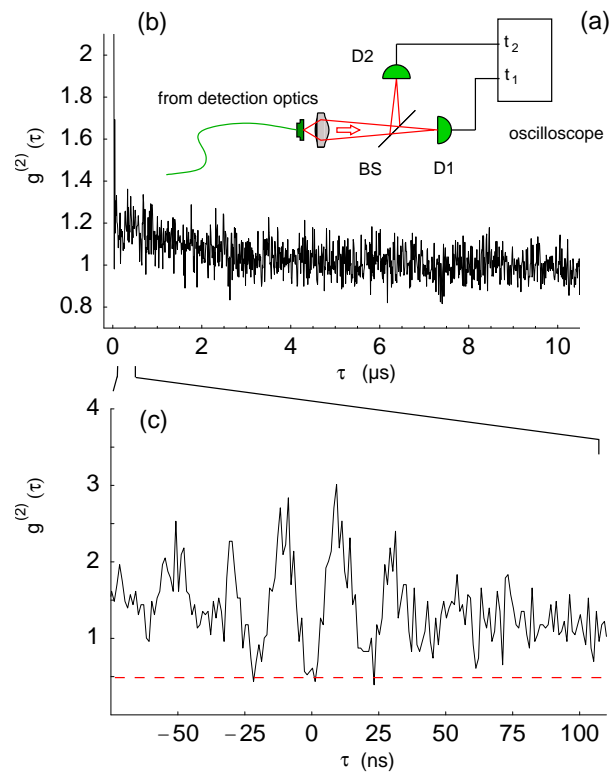}
\caption{(color online). (a) Hanbury-Brown-Twiss setup for the measurement of
the photon pair correlation function $g^{(2)}(\tau)$. The fluorescence light
is sent through a beam splitter (BS) onto two single photon detectors D1, D2
to record detection time differences $\tau = t_1 - t_2$. (b) On long
timescales, $g^{(2)}(\tau)$ shows a small bunching effect for $\mid\tau\mid
\le 2..3 \mu$s. (c) On short timescales, clear photon anti-bunching at
$\tau=0$ and oscillations due to Rabi flopping are observed. The dashed line
corresponds to accidental coincidences caused by the dark count rate of the
detectors. Experimental parameters: $I_{CL}=103$ mW/cm$^2$, $I_{RL}=12$
mW/cm$^2$, $\Delta_{CL}/2\pi=-31$ MHz, dipole trap depth $U=0.38$ mK.}
\label{Bild:hbtmessung}
\end{center}
\end{figure}

To assure that the upper fluorescence level corresponds to a single trapped
atom, we analyzed the non-classical properties of the emitted fluorescence
light. For this purpose the second order correlation function $g^{(2)}(\tau)$
was measured in an Hanbury-Brown-Twiss configuration with two detectors behind
a 50:50 beam splitter (inset (a) of Fig. \ref{Bild:hbtmessung}). The
differences of detection times $\tau = t_1 - t_2$ of photon pair events were
recorded in a storage oscilloscope with a conditional trigger mode. To
minimize background contributions, the coincidences are acquired only at
times, when the fluorescence exceeded a threshold of 1200 counts per second.

A normalized distribution of time differences $\tau$ is equivalent to the
second order correlation function as long as $\tau$ is much smaller than the
mean time difference between two detection events \cite{Reynaud83}. For
correct normalization of the measured $g^{(2)}(\tau)$ we divide the
coincidences in each time bin $\Delta\tau$ by $r_1 \times r_2 \times
\Delta\tau \times T$, where $r_1$ and $r_2$ are the mean count rates of the
two detectors, and $T$ is the total measurement time with an atom in the trap.

The resulting pair correlation function $g^{(2)}(\tau)$ for a trap depth
$U=0.38 \pm 0.04$ mK, a cooling laser intensity $I_{CL}\approx 103$ mW/cm$^2$,
and a detuning $\Delta_{CL}/2\pi$ of -31 MHz is shown in
Fig. \ref{Bild:hbtmessung}. On a $\mu$s timescale the correlation function
shows an exponential decay from the asymptotic value 1.24 around $\tau=0$ to
1.0 for large $\tau$ with a time constant of 1.8 $\mu$s. This effect can be
explained by the diffusive atomic motion in the intensity-modulated light
field of our three-dimensional cooling beam configuration and was already
studied in detail by Gomer et al. \cite{Meschede98} with a single atom
in a MOT.

On short timescales, most prominently we observe an uncorrected minimum value
$g^{(2)}(0)=0.52 \pm 0.14$ at zero delay ($\tau=0$). Taking into account
accidental coincidences due to the dark count rate of 300 $s^{-1}$ of each
detector, we derive a corrected minimum value $g^{(2)}_{corr}(0)=0.02 \pm
0.14$. Within our experimental errors this is compatible with perfect photon
anti-bunching of the emitted fluorescence light and therefore proves the
single atom character of our dipole trap. Furthermore, we observe the
signature of Rabi-oscillations due to the coherent interaction of the cooling
and repump laser field with the atomic hyperfine levels involved in the
excitation process. The oscillation frequency is in good agreement with a
simple two-level model \cite{Carmichael76} and the amplitude is damped out on
the expected timescale of the $5 ^{2}P_{3/2}$ excited state lifetime
$(=27$ns$)$.

The correlation function of a driven two-level atom shows its maximum value
$g^{(2)}_{max}=2$ for $\tau$ close to zero \cite{Carmichael76}. In contrast,
the background corrected correlations in Fig. \ref{Bild:hbttheory} show larger
oscillation amplitudes up to a maximum value of 5. This increase of the
oscillation amplitude - already known from experiments with single ions
\cite{Schubert92,Schubert95} - is a consequence of optical pumping among the
two hyperfine ground levels $F=1$ and $F=2$. To understand the consequences of
this effect on the second order correlation function in detail one has to take
into account the atomic level structure involved in the excitation process.

\subsection{Four-level model}

\begin{figure}[b]
\begin{center}
\setlength{\unitlength}{1cm}
\includegraphics[width=8cm]{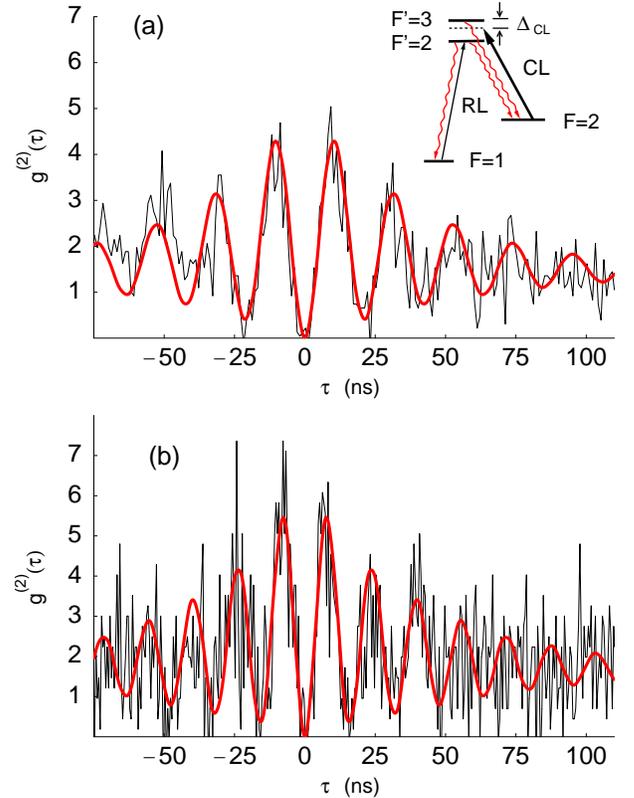}
\caption{(color online). Intensity correlation function $g^{(2)}(\tau)$
(background corrected) of the resonance fluorescence from a single $^{87}$Rb
atom in the dipole trap for two different trap depths and partial level scheme
of $^{87}$Rb (inset). Bold line: Numerical calculation. Thin line: measured
correlation function. Experimental parameters: $I_{CL}=103$ mW/cm$^2$,
$I_{RL}=12$ mW/cm$^2$, $\Delta_{CL}/2\pi=-31$ MHz, dipole trap depth (a)
$U=0.38$ mK, (b) $U=0.81$ mK.}
\label{Bild:hbttheory}
\end{center}
\end{figure}

For the fluorescence detection of a single atom in our dipole trap we use the
MOT cooling laser (CL), red detuned to the unperturbed hyperfine transition $5
^{2}S_{1/2}, F=2\rightarrow$ $5 ^{2}P_{3/2}, F'=3$ (inset of
Fig. \ref{Bild:hbttheory}) by $\Delta_{CL}=-4..5 \Gamma$ ($\Gamma=2\pi\times6$
MHz is the natural linewidth). To avoid optical pumping to the $5 ^{2}S_{1/2},
F=1$ hyperfine ground level we additionally shine in a repump laser (RL) on
resonance with the unperturbed hyperfine transition $5 ^{2}S_{1/2},
F=1\rightarrow$ $5 ^{2}P_{3/2}, F'=2$. Because the atom is stored in a dipole
trap the AC Stark-effect additionally shifts the cooling and repump light
fields out of resonance. This leads to significant atomic population in $F=1$
and therefore to a breakdown of the two-level assumption. 

For the following calculation we assume that the repump laser excites the
$F'=2$ level whereas the cooling laser can excite both hyperfine levels $F'=2$
and $F'=3$. These couplings are characterized by the Rabi frequencies
$\Omega_1$, $\Omega_2$ and $\Omega_3$, respectively. Because the three pairs
of counter-propagating circularly polarized cooling laser beams form an
intensity lattice in space and due to the finite kinetic energy of the atom
corresponding to a temperature of approximately 105 $\mu$K (see next section)
it is quite complicated to correctly describe the internal and external
dynamics of the atom in the dipole trap potential. In a classical picture the
atom oscillates in the trap potential with an amplitude of several optical
wavelengths. Hence, during this oscillatory movement the atom experiences
both, a changing intensity and polarization. This situation suggests to
simplify the internal atomic dynamics neglecting the Zeeman substructure of
the involved hyperfine levels (see Fig. \ref{Bild:hbttheory} (a), inset) and
to treat the exciting cooling and repump light fields as unpolarized with an
average intensity of six times the single beam intensity $I$.

The equation of motion for the atomic density matrix $\rho$ of this
system is given by
\begin{equation}
\dot{\rho} = \frac{-i}{\hbar}[H,\rho] + R. \label{liouville}
\end{equation}
In the rotating-wave approximation (RWA) the matrix representation of the
Hamiltonian $H$ -- describing the free atom and the interaction with the
repump laser field of angular frequency $\omega_{1}$ and the cooling laser
field of angular frequency $\omega_{2}$ -- in the basis of the bare atomic
states $|a\rangle$, $|b\rangle$, $|c\rangle$ and $|d\rangle$ corresponding to
the light-shifted hyperfine levels $F'=2, F=1, F=2$ and $F'=3$, respectively,
is given by:
\begin{equation}
H = \frac{-\hbar}{2} \left(
\begin{array}{cccc}
-2\omega_a & \Omega_{1} e^{-i t \omega_{1}} & \Omega_{2} e^{-i t \omega_{2}} 
& 0
\\  \Omega_{1} e^{i t \omega_{1}} & -2\omega_b & 0 & 0 
\\ \Omega_{2} e^{i t \omega_{2}} & 0 & -2\omega_c & \Omega_{3} 
e^{i t \omega_{2}} 
\\ 0 & 0 & \Omega_{3} e^{-i t \omega_{2}} & -2\omega_d
\end{array} \right).
\end{equation}
The relaxation term $R$ in the equation of motion (\ref{liouville}) represents
spontaneous decay \cite{Bloch53} from the excited hyperfine level $F'=3$ to
$F=2$ with a decay rate $\Gamma$ and from $F'=2$ to $F=2$ and $F=1$ with
$\Gamma/2$, according to the branching ratio of the respective hyperfine
transitions. In a matrix representation we obtain
\begin{equation}
R = \left(
\begin{array}{cccc}
  -\Gamma\rho_{aa}                        & -\frac{\Gamma}{2}\rho_{ab} &
  -\frac{\Gamma}{2}\rho_{ac}              & -\Gamma \rho_{ad} \\
  -\frac{\Gamma}{2}\rho_{ba}              & \frac{\Gamma}{2}\rho_{aa} & 
  0                                       & -\frac{\Gamma}{2} \rho_{bd} \\ 
  -\frac{\Gamma}{2}\rho_{ca}              & 0 &
  \frac{\Gamma}{2}\rho_{aa}+\Gamma\rho_{dd}  & -\frac{\Gamma}{2}\rho_{cd} \\ 
  -\Gamma \rho_{da}                                     & -\frac{\Gamma}{2} \rho_{db} & 
  -\frac{\Gamma}{2}\rho_{dc}              & -\Gamma\rho_{dd}
\end{array} \right),
\end{equation}
where energy relaxation from the $F=2$ to $F=1$ hyperfine ground level is
neglected.

The light field, scattered by the atom is described by the electric field
operators $\mathbf{E^{+}}$ and $\mathbf{E^{-}}$, and the two-photon
correlation function $g^{2}(\tau)$ is given, according to Glauber
\cite{Glauber63}, by
\begin{equation}
g^{(2)}(\tau) = \frac{\langle
\mathbf{E^{-}}(t)\mathbf{E^{-}}(t+\tau)\mathbf{E^{+}}(t+\tau)\mathbf{E^{+}}(t)\rangle}{{\langle \mathbf{E^{-}}(t) \mathbf{E^{+}}(t)
\rangle}^2}. \label{correlation}
\end{equation}
For almost monochromatic light fields and a small detection probability, this
function describes the conditional probability of detecting a photon at time
$t+\tau$, given the previous detection of another photon at time $t$,
normalized by the probability to detect statistically independent
photons. From the numerical solutions of the equation of motion
(\ref{liouville}) for the atomic density matrix $\rho$ we calculate
$g^{2}(\tau)$ with the help of the quantum regression theorem \cite{Lax66}
which relates the two-time expectation values in (\ref{correlation}) to
particular one-time expectation values and the initial conditions for the
density matrix \cite{Tannoudji98}. As we do not distinguish from which
hyperfine transition the first photon of a pair-event came from, the initial
condition $\rho(t=0)$ for the numerical solution of (\ref{liouville}) was
calculated from the steady-state solution $\rho(t=\infty)$. The resulting
correlation function is then given by the ratio of the excited state
populations at time $\tau$ and in the steady state ($\tau=\infty$)
\begin{equation}
g^{(2)}(\tau) = \frac{\rho_{aa}(\tau) +
\rho_{dd}(\tau)}{ \rho_{aa}(\infty) +
\rho_{dd}(\infty)}. \label{correlation_rho}
\end{equation}
For our experimental parameters we calculated the second order
correlation in (\ref{correlation_rho}) following the described procedure. To
include also the diffusive motion of the atom in the intensity modulated light
field of our cooling beam configuration, the resulting correlation function is
multiplied with $1 + A e^{-k \tau}$ \cite{Meschede98}, whereby the parameters
A and k have been determined from a fit to the measured correlation function
on the $\mu$s time scale.

Figure \ref{Bild:hbttheory} shows the measured, background corrected
correlation functions for two different dipole trap depths. Increasing the
dipole trap depth $U$ from 0.38 to 0.81 mK without changing the laser cooling
parameters enhances the effective detuning of the cooling laser to the
hyperfine transition $5 ^{2}S_{1/2}, F=2 \rightarrow 5 ^{2}P_{3/2}, F'=3$ due
to an increase of the AC Stark-shift of the respective atomic levels in the
far-off resonant dipole trap laser field. This effect gives rise to an
increase of the effective Rabi frequency from 47.5 MHz to 62.5 MHz and is
directly observed in the respective photon correlation function in
Fig. \ref{Bild:hbttheory}.

To summarize, within our experimental errors we find good agreement of the
calculated second-order correlation function with the measured
correlations. In contrast to a simple two-level model a four-level model is
required to correctly describe the observed oscillation amplitude of the
$g^{(2)}(\tau)$ function.

\section{Spectral Properties}

In the present experiment a single optically trapped atom is cooled by
three-dimensional polarization gradients in an optical molasses. This leads to
a final kinetic energy on the order of 100 $\mu$K \cite{Garraway00}. Due to
the motion of the atom in the confining potential the Doppler effect causes a
line broadening in the emitted fluorescence spectrum. Hence, a spectral
analysis of the emitted resonance fluorescence yields information about the
kinetic energy of the trapped atom.

For low excitation intensities the fluorescence spectrum of a two-level atom
exhibits an elastic peak centered at the incident laser frequency
$\omega_{L}$, while for higher intensities an inelastic component becomes
dominant, with contributions at the frequencies $\omega_{L}$ and $\omega_{L}
\pm \Omega_{0}$ \cite{Mollow69}, where $\Omega_{0}$ denotes the effective Rabi
frequency. This so-called ``Mollow triplet'' arises from the dynamical Stark
splitting of the two-level transition and has been observed in a number of
experiments, using low-density atomic beams \cite{Schuda74,Wu75,Hartig76} or a
single trapped and laser-cooled Ba$^+$ ion \cite{Stalgies96}. Surprisingly,
there are only few experimental investigations of the elastic scattering
process with a frequency distribution equal to the exciting laser. Sub-natural
line-widths were demonstrated with atomic beam experiments
\cite{Hartig76,Gibbs76}, atomic clouds in optical molasses
\cite{Westbrook90,Jessen92} and a single trapped and laser-cooled Mg$^+$ ion
\cite{Hoeffges97a,Hoeffges97b}.

For our laser cooling parameters the fluorescence spectrum is dominated by
elastic Rayleigh scattering \cite{Mollow69,Tannoudji98}. Hence, the emitted
fluorescence light exhibits the frequency distribution of the exciting laser
field (0.6 MHz FWHM) broadened by the Doppler effect. Position-dependent atomic
transition frequencies in the dipole trap due to the inhomogeneous AC-Stark
shift (caused by the finite kinetic energy) give no additional broadening,
because the spectrum of the elastically scattered fluorescence light is
determined only by the frequency distribution of the exciting light field and
not by the atomic transition frequencies.

\begin{figure}[t]
\includegraphics[width=8cm]{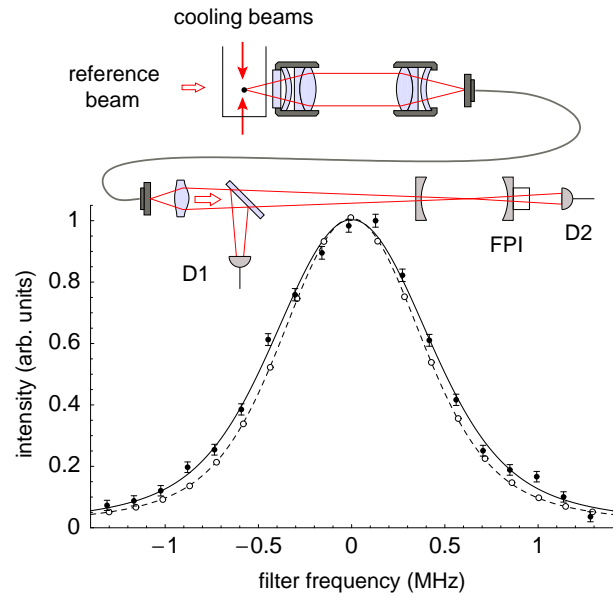}
\caption{(color online). Setup for the measurement of the resonance
  fluorescence spectrum of light scattered by a single $^{87}$Rb atom. Both,
  the atomic fluorescence and the laser light are analyzed alternately with
  the same scanning FPI. The spectra exhibit a width of 0.90$\pm$0.02 MHz and
  1.00$\pm$0.02 MHz (FWHM) for the excitation (-$\circ$-) and the fluorescence
  light (-$\bullet$-), respectively. Experimental parameters: $I_{CL}=87$
  mW/cm$^2$, $I_{RL}=12$ mW/cm$^2$, $\Delta_{CL}/2\pi=-19$ MHz, dipole trap
  depth $U=(0.62 \pm 0.06)$ mK.}
\label{pict:temperature}
\end{figure}

The scattered fluorescence spectrum is analyzed with a scanning Fabry-Perot
interferometer (FPI) with a frequency resolution of $0.45$ MHz (FWHM), a
transmission of $40\%$ and a finesse of 370. To measure the spectrum only at
times we trap a single atom, a part of the fluorescence light is monitored
separately with a reference APD (D1 in Fig. \ref{pict:temperature}). As the
broadening of the atomic emission spectrum due to the Doppler effect is small,
the instrumental function of the spectrometer and the exciting laser line
width have to be known accurately. In order to achieve this, we shine a
fraction of the exciting light (reference beam) into the collection optics
(see Fig. \ref{pict:temperature}). This way, both, reference and scattered
light are subject to the identical spectrometer instrumental function, whereby
the reference laser spectrum is also used to monitor length drifts of the
analyzing cavity. In the experiment, the spectrum of the reference beam and
the fluorescence light scattered by the atom were recorded alternately. After
each measurement a compensation of the cavity length drift was performed by
referencing the cavity frequency to the point of maximum transmission of the
reference laser.

With this procedure we obtained the two (normalized) data sets in
Fig. \ref{pict:temperature}. As expected, the resonance fluorescence spectrum
exhibits a ``sub-natural'' linewidth of $1.00 \pm 0.02$ MHz (FWHM) because the
elastic Rayleigh contribution dominates the scattering process. The exciting
laser light field exhibits a linewidth of $0.90\pm 0.02$ MHz (FWHM) which is
the convolution of the transmission function of the Fabry-Perot resonator with
the spectral width of the excitation laser. The depicted error bars reflect
the statistical error from the individual count rates of each data point. For
the reference laser spectrum this error is too small to be visible in this
graph.

For an atom at rest the resonance fluorescence spectrum shows the same
linewidth as the exciting light field. Any finite kinetic energy distribution
of the atom will lead to a broadening of the atomic emission spectrum and
therefore can be used for the determination of the atomic ``temperature''. To
extract the mean kinetic energy from the measured spectra in Figure
\ref{pict:temperature}, we assume that the atom is subject to the same
stationary Gaussian velocity distribution in all directions. This assumption
is justified because the atom is expected to occupy on average up to 100
motional modes of the dipole trap potential \cite{Garraway00}. Therefore the
atomic motion can be considered classical and the energy distribution is given
by the Boltzmann statistics, leading to a thermal velocity distribution.

According to this assumption we convolve a Gaussian velocity distribution with
the measured reference laser line profile. The resulting function is fitted to
the data points of the fluorescence spectrum with the variance of the Gaussian
profile being the only free fit parameter \footnote{This procedure is
justified because the statistical error on the data points of the reference
laser spectrum is much smaller than the error on the fluorescence data.}. From
the fitted variance we directly obtain the mean kinetic energy $E_{kin}$ of a
single atom in the dipole trap
\begin{equation} \label{equ:temperatureRESULT}
E_{kin}=\frac{1}{2}m\langle \Delta v^2\rangle=(105 \pm
24)^{+14}_{-17}\mbox{ }\mu \mbox{K}\cdot k_B,
\end{equation}
with a statistical error of $\pm 24$ $\mu$K. Here $k_B$ denotes the Boltzmann
constant, $m$ the atomic mass and $\langle \Delta v^2 \rangle$ the mean
quadratic velocity.

The calculation of the mean kinetic energy contains a systematic error because
the cooling beams have different angles relative to the axis defined by the
dipole trap and the detection optics. The overall Doppler broadening of the
elastically scattered fluorescence light depends on these angles. Because the
relative intensity of these beams is not known exactly, a systematic error
occurs. In order to estimate an upper bound for this error we assume that the
atoms scatters light only from the beams which would give the highest or
lowest velocities, respectively. From this estimation we obtain the last two
error bounds in (\ref{equ:temperatureRESULT}). Within the experimental errors,
the measured temperature is equal to or smaller than the Doppler temperature
of $^{87}$Rb (146 $\mu$K).

\section{\label{sec:level1} Summary}

We have studied the non-classical properties of fluorescence light scattered
by a single optically trapped $^{87}$Rb atom. For this purpose, we have set up
an HBT experiment and measured the second order correlation function of the
detected fluorescence light. The measured two-photon correlation function
exhibits strong photon antibunching verifying the presence of a single trapped
atom. Due to inelastic two-body collisions which are present during the
loading stage of the dipole trap and the small trap volume \cite{Schlosser01},
only a single atom per time was trapped. Furthermore, the measured second
order correlation function shows the internal and external dynamics of the
atomic hyperfine levels involved in the excitation process. An atomic
four-level model was developed and its predictions were compared with the
measured second order correlation functions. Within the experimental errors we
find good agreement of the calculated predictions with the measured data.

In addition, the spectrum of the emitted resonance fluorescence was
measured. We find, that the atom-light interaction is dominated by elastic
Rayleigh scattering. Due to the Doppler effect, we observed an additional
broadening of the atomic fluorescence spectrum. From this we determined the
mean kinetic energy of the trapped atom corresponding to a temperature of 105
$\mu$K.

This simple single-atom trap is a promising tool for the effective generation
of narrow-band single photons \cite{Grangier05} and for the realization of a
quantum memory for light \cite{Kuzmich04}. Furthermore, our setup can also be
used for the generation of entanglement between the spin state of a single
$^{87}$Rb atom and the polarization state of a spontaneously emitted single
photon \cite{Weber05}. This kind of atom-photon entanglement will close the
link between atoms and photons in quantum information applications and opens
the possibility to entangle atoms at large distances, well suited for a
loophole-free test of Bell's inequality \cite{Karen,Simon03,Weber05}.

\end{document}